\documentstyle[aps,eqsecnum,epsfig]{revtex}

\begin{document}
\draft
\title{Weak violation of universality for Polyelectrolyte Chains: \\
                        Variational Theory and Simulations}

\author{Gabriele Migliorini$^{1}$, Vakhtang G. Rostiashvili$^{1}$, 
and Thomas A.
Vilgis$^{1,2}$}
\address{$^1$ Max Planck Institute for Polymer Research\\
10 Ackermannweg, 55128 Mainz, Germany.\\
$^2$ Laboratoire Europ\'een Associ\'e, Institute Charles Sadron\\
6 rue Boussingault, 67083 Strasbourg Cedex, France.}

\date{\today}

\maketitle

\begin{abstract}
  
  A variational approach is considered to calculate the free energy
  and the conformational properties of a polyelectrolyte chain in $d$
  dimensions. We consider in detail the case of pure Coulombic interactions
  between the monomers, when screening is not present, in order to compute the
  end-to-end distance and the asymptotic properties of the chain as a function
  of the polymer chain length $N$.  We find $R \simeq N^{\nu}(\log
  N)^{\gamma}$ where $\nu = \frac{3}{\lambda+2}$ and $\lambda$ is the exponent
  which characterize the long-range interaction $U \propto 1/r^{\lambda}$. The
  exponent $\gamma$ is shown to be non-universal, depending on the strength of
  the Coulomb interaction. We check our findings, by a direct numerical
  minimization of the variational energy for chains of increasing size
  $2^4<N<2^{15}$. The electrostatic blob picture, expected for small enough
  values of the interaction strength, is quantitatively described by the
  variational approach. We perform a Monte Carlo simulation for chains of
  length $2^4<N<2^{10}$.  The non universal behavior of the exponent $
  \gamma$ previously derived within the variational  method, is also confirmed 
  by the simulation results. Non-universal behavior is
  found for a polyelectrolyte chain in $d=3$ dimension. Particular attention
  is devoted to the homopolymer chain problem, when short range contact
  interactions are present.

\end{abstract}


\section{Introduction}

Charged polymers play a central role in theory and experiments.
Polyelectrolytes are macromolecules containing ionizable groups. They
dissociate in water to form charged groups and low-molecular-weight counter
ions.  It is important to remark that interactions between different polyions
in a $many~chain~system~$ are of crucial importance for any realistic system
with a low but finite overlap concentration. The problem of the $single~chain$
conformation is the one of the essential questions for polyelectrolyte
solutions and has been under investigation in the recent years.  This will be
the topic of the present paper as well, but in a more general context. As
common to most of the recent investigation, we consider the limit of infinite
dilution and zero salt concentration for polyelectrolytes.  In this limit a
polyelectrolyte can be represented as a connected sequence of charged and
uncharged monomers in a ``dielectric~vacuum'' that constitutes the solvent.
>From a scaling point of view, the properties of single polyelectrolyte chains,
despite the long time debates are still not completely clear.  Indeed, for the
neutral polymer chain system, where a clear understanding of the scaling
properties has been achieved \cite{Gennes}, the details of the local chain
structure are not essential, while in the polyelectrolyte chain problem they
could be important \cite{Gennes2,Joanny}.  In particular the long range nature
of the Coulomb interaction couples via the presence of the solvent, the short
and long length scales of the system . Thus a scaling approach to
polyelectrolytes results to be more involved than for neutral chains.  The
interaction between monomers, counterions and solvent molecules leads to the
important phenomenon of counter ion condensation \cite{Manning} and a
realistic description should take into account all the different characters of
this scenario.  On the other hand, a minimal  or bare essential description
of the system is required in order to achieve a quantitative description of
the single polyelectrolyte chain problem.

In this paper we will present a generalized form of des Cloizeaux Gaussian
variational method \cite{Cl,Cloizeaux} to study the scaling behavior and
local properties of polyelectrolytes.  Variational techniques have been shown
to be well suited to describe a self-interacting polymer with long range
repulsion \cite{Bouchaud}.  Scaling properties of the end-to-end distance of
the chain have been discussed for the long - ranged intra-molecular
interaction $1/r^{\lambda}$ at an arbitrary spatial dimensionality $d$.  A
recent numerical implementation of such a variational technique, for the case
of Coulombic interaction (when $\lambda = d - 2$), seems to be promising as
well \cite{Bratko}.  On the other hand, the validity of the variational
approach for short range interacting homopolymer chains must be questioned
\cite{Bouchaud,Bratko,Allegra1} for natural and well known
reasons.  The most studied example
in the context of polymer physics is the problem of the self avoiding chain.
It was proven  that the problem shows universality in the limit of infinitely
long chains \cite{Gennes} and can be treated by field theoretic  methods
common to phase transitions. Indeed the renormalization group (RG)
theory manifests these statements and develops systematic methods to 
compute the scaling exponents, e.g. for the size of the chain  
\cite{Cloizeaux}. The variational technique provides 
an alternative approach  since most of the approximations can be
controlled.  
Although, it is well known from elementary statistical mechanics that 
the variational technique cannot provide such powerful tool due to 
their mean - field character. Therefore this technique is expected 
to work in the mean field limit, i.e. at large dimensions 
(at the upper critical dimension) or for long range potentials.  
Thus it is well investigated and generally believed
that the Gaussian variational approach is better suited for long-range
interacting chains.  

Moreover, it has been suggested that in the study of
random heteropolymer \cite{SG1,SG2} and directed polymer in a random media
\cite{MP}, a variational approach in replica space can be also successfully
applied. We postpone these issues for future investigations. Here we will
mainly concentrate on the pure long-range and short-range homopolymer problems
within the Gaussian variational technique.

In section II we introduce the Gaussian polymer model and calculate the
variational free energy using a general Gaussian trial Hamiltonian in the
spirit of des Cloizeaux \cite{Cl,Cloizeaux} and reference \cite{NO}.  We then
critically review the asymptotic analysis given originally in reference
\cite{Cl} and suggest  a new one  which leads to the Flory exponents
corrected by the presence of logarithmic factors.  In section III we describe
a numerical technique in order to solve the Euler equations corresponding to
the variational principle, by means of a new algorithm specially devised for
this problem.  An independent algorithm, that simply minimizes the variational
free energy is shown to be quantitatively equivalent to the direct solution of
the Euler equations. The numerical solution of the Euler equations is then
carefully analyzed for polymer chains of increasing size. We show here that
our new asymptotic analysis is in quantitative agreement with our numerical
findings.  We discuss connections with the short range homopolymer
problem 
within the Gaussian Variational Principle and results for the end-to-end
distance are obtained. In section IV Monte Carlo simulation results, obtained
by means of a pivot algorithm, are presented, showing quantitative agreement
with our variational analysis.  The case of a self-interacting homopolymer
chain, when an attractive two-body interaction between the monomers is
considered together with a three body repulsive term, can also be studied
within our variational approach.  Preliminary calculations shows that the
globular formation, expected from simple phenomenological arguments and due to
the balancing between the two body and three body interactions, takes place.
The instability of the globule state due to  a Coulombic repulsion
between the monomers  is also  described by our variational principle and will be presented
elsewhere \cite{Migliorini}.  Conclusions and further prospects are summarized
in section V.

\section{The Gaussian Variational principle}

\subsection{Basic equations}

First we start from  a well known example for a Gaussian variational approach
suitable for the investigations presented below.  We introduce a simple
problem to compute the free energy of a self-interacting polymer chain with
Coulombic repulsion between the monomers, based on a discrete representation
of a chain of $N$ monomers. The method relies on the well known variational
principle, where a Gaussian trial probability  is proportional to the
exponential of a quadratic form of the monomer coordinates $\vec{r_j}$,
\begin{equation}
P_V(\vec{r_1},..,\vec{r_N}) = Z_V^{-1} \exp \{
-H_0(\vec{r_1},..,\vec{r_N}) \}.\label{trial}
\end{equation}
In eq.(\ref{trial}) the trial Hamiltonian  $H_0$ is given by
\begin{equation}
H_0(\vec{r_1},..,\vec{r_N}) = \frac{d}{2} \sum_{i=1}^{N} \sum_{j=1}^{N}
G_N(i-j) (\vec{r_i} - \vec{r_j})^2
\label{Gauss}
\end{equation}
and $Z_V$ is a normalization constant determined by the condition
\begin{equation}
  \int P_V(\vec{r_1},..,\vec{r_N})d^d \vec{r_1}\dots d^d \vec{r_N} =1.
\end{equation}
Up to this point different forms of the Gaussian variational 
principle could be considered. 
We decided here to consider the variational form originally proposed by
des Cloizeaux \cite{Cl} in the context of the self-avoiding homopolymer
problem.  The main essential features of the Gaussian variational principle
can be obtained within this choice of the variational form according to
equation (\ref{Gauss}).  More sophisticated forms of the variational kernel
including a classical path $\vec{r_0}$ and eventual anisotropy along parallel
and perpendicular directions to the polymer backbone, can be also  included \cite{NO}.  By the appropriate
choice of a periodic boundary conditions for the chain
the quadratic form $H_0$ can be diagonalized by introducing cyclic coordinates
(Rouse modes) for the monomer positions  
\begin{eqnarray}
\vec{ \rho_q} &=& N^{-\frac{1}{2}} \sum_{j=1}^N \exp [~ {\text i} 2 \pi j q/N ]
\vec{r_j} \nonumber \\
~~~\vec{r_j} &=& N^{-\frac{1}{2}} \sum_{q=1}^N \exp [- {\text i} 2 \pi j q/N ]
\vec{ \rho_q}.
\end{eqnarray}
Indeed, due to cyclic invariance
the Cartesian components of $\vec{ \rho}_q$ satisfy
\begin{equation}
\langle \rho^{(j)}_q \rho^{(j')}_{q'} \rangle = \frac{1}{d}
\delta_{jj'} \delta_{q q'} g_N^{-1}(2 \pi q/N),
\end{equation}
where   $g_N^{-1}(2 \pi q/N)$ is positive and
discontinuous function of $k=2 \pi q/N$  and is related to the
Gaussian propagator $G$ of equation (\ref{Gauss}) via the following
expression
\begin{equation}
g_N(2 \pi q/N) = \sum_{n =1}^{N-1} G_N(n) [1-\cos(2 \pi qn/N)].
\end{equation}
The variational free energy for a uniformly charged polymer of
polymerization index $N$ in a $d$ dimensional space can be computed
according to the usual Feynman inequality
\begin{equation}
F \le F_V = \langle H - H_0 \rangle_{0} +F_0,
\label{Inequal}
\end{equation}
where $F_0$ is the free energy of the Gaussian model defined by
equation (\ref{Gauss}). 
In equation (\ref{Inequal}) $H$ represents the full Hamiltonian of a
charged polymer chain with a pure long - range  form
for the monomer-to-monomer interaction,
\begin{equation}
 V(| \vec{r_i} - \vec{r_j} |) = | \vec{r_i} - \vec{r_j}| ^{ - \lambda},
\label{potential}
\end{equation}
where $\lambda < d$. The expression for the Hamiltonian reads
\begin{equation}
H = \frac{d}{2b^2} \sum_{i=1}^N (\vec{r_{i+1}} - \vec{r_i})^2 +
\beta/2 \sum_{i=1}^N \sum_{j=1}^N V(| \vec{r_i} - \vec{r_j} |),
\end{equation}
where $\beta = l_{\rm B}^{\lambda}/A^2$ depends on the Bjerrum length
$l_{\rm B} =(e^2/4 \pi \epsilon k_B T)$ and where the distance
between charged monomers along the chain is given by $A$ in units of the
Kuhn segment $b$. $A$ defines thus the charge fraction $f=1/A$.

It is straightforward to derive the following expression
for the variational free energy $F_V$ in the limit of infinitely long
chain $N \rightarrow \infty $,
\begin{eqnarray}
F_V[g(k)]/N=\frac{d}{2} b(1) &+& \beta
\left(\frac{d}{2}\right)^{\frac{\lambda}{2}}
\frac{\Gamma\left(\frac{d - \lambda}{2}\right)}
{\Gamma\left(\frac{d}{2}\right)}\sum_{n=1}^{ \infty}
[b(n)]^{- \frac{\lambda}{2}}+ \nonumber \\
\frac{d}{4 \pi} \int_{ - \pi}^{ + \pi} \log g(k) dk
 &-& \frac{d}{2} [ 1+ \log (2 \pi/d) ],
\label{Euler4}
\end{eqnarray}
where
\begin{equation}
b_N(i-j)= \frac{1}{N} \int_{- \pi}^{ +\pi} \frac{(1-
  \cos((i-j)k))}{g(k)}dk,
\label{Euler5}
\end{equation}
\begin{equation}
b_N(i-j) \equiv \langle (\vec{r_i} - \vec{r_j})^2 \rangle,
\label{Square}
\end{equation}
and 
\begin{equation}
\lim_{N \rightarrow \infty}g_N(k_N) \equiv g(k),~~g(k) \ge 0.~~~~~~~~~
\label{Euler1}
\end{equation}
The function $g(k)$ satisfies the symmetry requirements
\begin{equation}
g(k)=g(-k),~~g(k+2 \pi)=g(k).
\label{Euler3}
\end {equation}
Minimization of the variational free energy $F_V$ yields
\begin{eqnarray}
\label{Euler2}
g(k) = 1- \cos(k) &-& \beta \left(\frac{\lambda}{2} \right)
\left(\frac{d}{2} \right)^{ \frac{\lambda}{2}}
\frac{\Gamma \left(\frac{d- \lambda}{2} \right)}
{\Gamma \left(\frac{d}{2} \right)} \nonumber \\
\times \sum_{n=1}^{ \infty} (1 &-& \cos(nk)) [b(n)]^{-
\frac{\lambda+2}{2}}.
\end{eqnarray}
Equations (\ref{Euler4})-(\ref{Euler2}) represent the generalization
of the Gaussian variational principle for the long-range potential
(\ref{potential}) which was introduced first in reference
\cite{Bouchaud}.
For the pure Coulombic case $ \lambda=d-2$, comparing the expression
for the variational free energy (\ref{Euler4}) with the correspondent
expression for the short range interaction problem \cite{Cl},
\begin{eqnarray}
F_V[g(k)]/N=\frac{d}{2} b(1) &+& w
\sum_{n=1}^{ \infty} [b(n)]^{- \frac{d}{2}}+ \nonumber \\
\frac{d}{4 \pi} \int_{ - \pi}^{ + \pi} \log g(k) dk
 &-& \frac{d}{2} [ 1+ \log (2 \pi/d) ],
\label{Euler4bis}
\end{eqnarray}
leads to the following conclusion which  holds within the Gaussian Variational Principle. 
All universal
exponents for the Coulombic case are related to the corresponding
exponents for the short-range interaction case through the dimensional
shift 
\begin{equation}
d \rightarrow d-2 .
\label{shift}
\end{equation}
The result of minimization of equation (\ref{Euler4bis}) reads
\begin{equation}
g(k) = 1 - \cos(k) - w
\sum_{n=1}^{ \infty} (1 - \cos(nk)) [b(n)]^{- \frac{d+2}{2}},
\label{Euler2bis}
\end{equation}
which is of the same form like  equation (\ref{Euler2}).

Before moving to the next section, where a careful numerical
implementation of the Euler equations (\ref{Euler5})-(\ref{Euler2}) for
polymer chains of increasing length $N$ is presented, we revise
the asymptotic analysis of the Euler equations in the limit of small
$ k \rightarrow 0 $,which was  originally given by des Cloizeaux \cite{Cl}.
The behavior of $g(k)$ at small $k$ determines  the asymptotic behavior
of $b(n)$ for $ n >> 1$  via equation (\ref{Euler5}).
Let us assume first that in this limit  $g(k)$ has the following form

\begin{equation}
g(k) \simeq g |k|^{1+ 2 \nu},~\frac{1}{2}< \nu <1,
\label{ansatz}
\end{equation}
then the  correlation function $b(n)$, reads
\begin{equation}
b(n) \simeq b n^{ 2 \nu}.
\label{assympt}
\end{equation}
We will now proceed to analyze the Euler
equations in the scaling limit of small $k$ values
within this simple ansatz.

\subsection{The asymptotic analysis - neutral chains}

To demonstrate several problems within 
the Gaussian variational principle
we are going to rewise in 
this section  the asymptotic analysis of the Euler
equation  (\ref{Euler2bis}) for neutral chains,
which was originally given in
the context of the self avoiding chain problem \cite{Cl,Cloizeaux},
and later extended to self interacting polymer chains with long-ranged
Coulomb like interactions \cite{Bouchaud}.
It has been suggested thatÄ due to the long ranged nature of the
interactions between the monomersÄ the Gaussian Variational Principle
is particularly well
suited to describe the asymptotic behavior of polyelectrolyte
chains. Special attention has also been devoted \cite{Bratko,NO,Sodeberg}
to the case of screened Coulombic interaction,
within the Debye H\"uckel approximation.

Let us consider the interaction term
\begin{equation}
I(k)=Ä w \sum_{n=1}^{ \infty} (1 - \cos(nk)) [b(n)]^{- \frac{d+2}{2}}
\label{summa}.
\end{equation}
We can obtain the asymptotic expression for $I(k)$ in the
$k \rightarrow 0$ limit, substituting the series with an integral
\cite{Cl} and  assuming that the dominant contribution would correspond
to large values of $n$. The result can be written in the form
\begin{equation}
I(k) \simeq - w \frac { \pi b^{ -\frac{d+2}{2}} }{ 2 \Gamma( \delta)
\cos( \frac{ \pi}{2} \delta )} k^{ \delta -1} +
w b^{ - \frac{d+2}{2}} \zeta( \delta-2) \frac{k^2}{2} ,\label{expansion}
\end{equation}
where $ \delta = \nu (d+2)$.

An alternative derivation can be obtained in the asymptotic regime
$n >>1$Ä evaluating exactly the series
\begin{equation}
S( \delta ,z) = \sum_{n=1}^{\infty}
\frac{ z^{ n}}{n^{ \delta}}Ä ,
\label{series0}
\end{equation}
atÄ $z=\exp(ik)$ . The expliciteÄ calculations are given in Appendix
where it is stressed that we  should be distinguish   between noninteger
and integer $\delta$. WeÄ apply these results for calculation of
eq. (\ref{summa}) where $b(n)$ is given by eq.(\ref{assympt}).In this
case weÄ find
\begin{equation}
I(k) \simeq w b^{- \frac{d+2}{2}} \left[ S(\delta, z=1)-{\text Re}~S(\delta,
Ä z) \right] ,
\end{equation}
where $ \delta = \nu (d+2)$.
Considering non-integer values of the exponentÄ $\delta$Ä (this
assumption will be
checked afterwards)
eq. (\ref{seriesexp})Ä yields
\begin{eqnarray}
\label{series}
I(k) &=& wÄ b ^{- \frac{d+2}{2}}\Bigl[- \frac { \pi }{ 2 \Gamma( \delta)
\cos( \frac{ \pi}{2} \delta )} k^{ \delta -1} 
-
\sum_{p=1}^{ \infty} \zeta( \delta - 2p) (-1)^p \frac{ (k)^{ 2p} }{(2p)!}\Bigr].
\end{eqnarray}
This equation generalizes
eq. (\ref{expansion})ÄÄ ( see alsoÄ eq. (B.9)
of ref. \cite{Cl} ). The Euler equation (\ref{Euler2bis}) together with the
expansion (\ref{series}) gives, in the $k \rightarrow 0$ limit
\begin{eqnarray}
gk^{1+ 2 \nu} = \frac{1}{2}k^2-w b^{ - \frac{d+2}{2}}
\zeta( \delta-2) \frac{k^2}{2} 
+
w \frac { \pi b^{ -\frac{d+2}{2}} }{ 2 \Gamma( \delta)
\cos( \frac{ \pi}{2} \delta )} k^{ \delta -1}+ O(k^4) ,
\label{asymptotic}
\end{eqnarray}
where $ \delta = \nu (d+2)$.

Let us impose first the balancing conditions
in the spirit of des Cloizeaux: (i) the first two terms on the r.h.s.
balance each other, (ii) the term on the l.h.s. and the third term
on the r.h.s. are also balanced. Then we immediately obtain
\begin{equation}
\nu = \frac{2}{d}~,
\label{wrongexpo}
\end{equation}
as a valid exponent, but together with the additional conditions
\begin{eqnarray}
\label{balance0}
~&w& b^{ -\frac{d+2}{2}} \zeta( \delta-2)=1,Ä \\
~&g& = \frac{ w \pi b^{ -\frac{d+2}{2}} }
{ 2 \Gamma (\delta) \cos( \frac{ \pi}{2} \delta )},
\label{balance}
\end{eqnarray}
contained in the above equations, the physical interpretation
becomes, however, more complicated.
Moreover, we must add to these equations the general relation between $g$ and
$b$, which originates from equation (\ref{Euler5})
\begin{equation}
b = \frac{1}{ 2 g \Gamma (1+ 2 \nu) \sin( \pi \nu )} .
\label{balance2}
\end{equation}
Equation (\ref{wrongexpo}) is thus the
well known result for the exponent
$ \nu$, whereas equations (\ref{balance0})-(\ref{balance2}) set up three
equations for the twoÄ unknowns $b$,$~g$.
Equations (\ref{balance0})-(\ref{balance2}) determines not
only the amplitudes $g$ and $b$ but also {\it fixes} the value of 
the interaction
parameter $w$ itself. This delicate point, (see equation (IV.52) of
reference \cite{Cl}) has already been discussed by Allegra
\cite{Allegra1}.
This is contradictory in the sense that generally
the interaction parameter $w$ is an arbitrary quantity. The proper balance
condition must consist of a single equation, whereas the conditions
(\ref{balance0})-(\ref{balance}), derived in the asymptotic analysis
of des Cloizeaux, overconstrains the parameters involved.
Besides thatÄ the parameter $\delta =2(d+2)/d$
is only non-integer for $d=3$ and is integer for $d=1,2,4$ (where
the Flory scaling produces exact results). This indicates
that the des Cloizeaux solution can not be valid in this later case
since it relies on the expansion (\ref{series}) which (see Appendix)
is only valid for non-integer values of $\delta$.

In order to avoid the overconstraints we  try to impose the Flory
balancing conditions : (i) the elastic term,
which is proportional toÄ $k^2$, is balanced by the interaction term
eq.(\ref{summa}), 
(ii) the term on the l.h.s. (or the entropic term) is negligible
compare to the elastic and interaction terms on the r.h.s. Obviously
the Flory balancing conditions have one constraint less. In this
case for $S(\delta, z )$ we must use eq.(\ref{asym}) where $\delta= m
= \nu(d+2)=3$ and $z=\exp ( ik )$. This immediately leads to 
the well known Flory exponent
$\nu_{\rm F} = 3/(d + 2 )$ . Then the interaction term (\ref{summa}) becomes
\begin{eqnarray}
&&I(k) =Ä \lim_{k \rightarrow 0}w b^{-\frac{d+2}{2}}\left[ S(m = 3 , z=1)-Re~S(m = 3, z) \right] =
\nonumber\\
&&w b^{-\frac{d+2}{2}} \sum_{n=1}^{ \infty} \frac{1- \cos(nk) }{n^{ 3 }} \simeqÄw b^{-\frac{d+2}{2}}Ä \left[ \psi(3) - \psi(1) - \log k \right]\frac{ k^2 }{2 }+ {\cal O}(k^4).
\label{asym2}
\end{eqnarray}
The resulting Euler equation takes the form
\begin{eqnarray}
&g& k^{1+ 2 \nu_{\rm F}} = \frac{1}{2}k^2-w b^{ - \frac{d+2}{2}}\left[ \psi(3) - \psi(1) - \log k \right]\frac{ k^2 }{2 }
+ O(k^4) .
\label{asymptotic2}
\end{eqnarray}
Eq.(\ref{asymptotic2}) can be also obtained directly from
eq.(\ref{asymptotic}) if we assume $\delta =3 + \epsilon$,Ä where
$\epsilon \to 0$. Then the poles which have its origin from the second and
third terms on the r. h. s. are canceledÄ (see e.g. Appendix and sec. 1.11 in
ref.\cite{erdelyi}) but the term $k^{2}\log k$ appears.
TheÄ presence of
the $k^2 \log k$ term in equation
(\ref{asymptotic2}) leads to the conclusionÄ that
the pure power law ansatz $g(k) \propto k^{1+2 \nu}$ (or $b(n)
\propto n^{2 \nu}$)
does not solve in the asymptotic limit
$k \rightarrow 0~$Ä the Euler equation (\ref{Euler2bis}) once the
exponent $\delta=(d+2) \nu$ is assumed to be integer. In fact, the
direct numerical solution of the Euler equations (\ref{Euler2}) and
(\ref{Euler2bis}) presented in the next section does not show a
Äpure
power law behavior in the asymptotic regime $k << 1$.
We also see that the term $k^2 \log k$ naturally appears in the
previous expressions for the series $S( \delta,k)$, which suggests to
consider
logarithmic corrections in the initial ansatz for $b(n)$. This will
allow to stay within a Flory-like solution
(corresponding to the integer value $\delta = \nu (d+2)=3$) and also
to compute explicitly the exponent that characterize the logarithmic
correction to the pure power law scaling behavior.
Moreover the problem of having three equations for the two
parameters $g$ and $b$, which invalidates the
asymptotic analysis of des Cloizeaux, is naturally solved within the
new ansatz we proposeÄ in the next subsection. The value of $w$ is
no longer constrained to a certain value
as it should be the case
in an ideal experiment where it is
allowed to tune the Coulombic interaction (or the quality of the
solventÄ for the self avoidingÄ problem).

\subsection{Corrections to scaling}
As it was suggested we will stay within the Flory balancing condition
but try the next   asymptotic ansatz for the correlation function
(\ref{Square})
\begin{equation}
b(n) \simeq b_0 n^{ 2 \nu} (\log n)^{ 2 \gamma},
\label{newansatz}
\end{equation}
where $ \gamma$ is the exponent that characterize the logarithmic
correction to scaling.
Substituting the ansatz (\ref{newansatz}) in the Euler equation
(\ref{Euler2bis}) gives for the interaction term
\begin{equation}
I(k)=w b_0^{- \frac{d+2}{2} } \sum_{ n=1}^{ \infty}
\frac{ 1- \cos(nk) } {n^{ \nu (d+2)} [ \log n ]^{ \gamma (d+2)}}.
\end{equation}

The largest contribution to this  series at small $k$  comes from the
large $n$ limit. We can substitute the sum with an integral 
(for the power law ansatz, this is explained and justified in the 
Appendix)  to obtain
\begin{equation}
I(k) \simeq  w b_0^{- \frac{d+2}{2} }  \int_{\bar{n}}^{ \infty} dn \frac{ 1- \cos(nk) } {n^{ \nu (d+2)}
 [ \log n ]^{ \gamma (d+2)}},
\label{int}
\end{equation}
where $\bar{n} $ is a cutoff we introduce to regularize the integral.
Equation (\ref{int}), after substituting $n=x/k$ reads
\begin{equation}
I(k) \simeq k^{ \nu (d+2) -1} w b_0^{- \frac{d+2}{2} }  \int_{\bar{x}}^{ \infty} dx 
\frac{ 1- \cos(x) } {x^{ \nu (d+2)} [  \log x - \log k ]^{ \gamma (d+2)}},
\label{int1}
\end{equation}
where, because of  the Flory balancing conditions ($I(k)\propto
k^2$),  
we should impose $ \nu(d+2) = 3$. 
Let us consider a value $t$ such that at $ x < t$ it is ensured that
$1 - \cos x \simeq x^2/2$ holds. In this case the integral  in
equation (\ref{int1}) becomes
\begin{eqnarray}
J(k) &=&  \int_{\bar{x}}^{ \infty} dx \frac{ 1- \cos(x) }
{x^{3} [  \log x - \log k ]^{ \gamma (d+2)}} \nonumber\\
&\simeq& \frac{1}{2}\int_{k \bar{n}}^{ t } dx
\frac{1}{x[  \log x - \log k ]^{ \gamma (d+2)}} \nonumber \\
&+& \int_{t}^{ \infty } dx
\frac{1- \cos(x) }{x^3[  \log x - \log k ]^{ \gamma (d+2)}}=J_1+J_2.
\end{eqnarray}
At $k \rightarrow 0$ $J_2(k) \rightarrow 0 $, while the integral
$J_1(k)$ defined above can be evaluated as
\begin{equation}
J_{1}(k)= \frac{1}{2[ \gamma(d+2)-1 ]}
[ [ \log(\bar{n})]^{ 1- \gamma (d+2)}],
\end{equation}
provided that $\gamma (d+2) > 1$ and $\bar{n} >1$.
We conclude that the Euler equation (\ref{Euler2bis}) is properly
regularized by the logarithmic ansatz (\ref{newansatz}) and the cutoff
parameter $ \bar{n}>1$. The balance in
equation (\ref{Euler2bis}), between the elastic energy term and the
interaction term, gives
\begin{eqnarray}
\label{inequality0}
~~~\nu &=& \frac{3}{d+2}, \\
~~~\gamma &>& \frac{1}{d+2}.
\label{inequality}
\end{eqnarray}

\subsection{Correction to scaling - charged chains}

For the Coulombic case, considering the dimensional shift rule 
(\ref{shift}), we obtain
\begin{eqnarray}
\label{inequality1}
\nu &=& \frac{3}{d},\label{florexp} \\
\gamma &>& \frac{1}{d}~.
\label{inequality2}
\end{eqnarray}
The inequality we just derived will be verified numerically 
in the next section.
Values for the exponent $\gamma$ will be obtained via the numerical
solution of the Euler equations (\ref{Euler4})-(\ref{Euler2}) and
via Monte Carlo simulations. 
Within  our asymptotic analysis, we can
obtain an expression for the exponent $\gamma$ that depends on the
interaction strength. In particular by balancing the prefactors of the proper
terms we have
\begin{equation}
\gamma \simeq  \frac{1}{d} \left(1+ \frac{w}{b_{0}(w) ^{d/2}} \right),
\end{equation}
where we also took into account that the amplitude $b_{0}$ can depend
from $w$.
The exponent $\gamma$ depends on the interaction 
strength, being non universal. This will be tested 
against simulations and numerical minimization of the variational 
free energy (\ref{Euler4}).

It was suggested \cite{Joanny} that above a certain minimum
charge fraction $f \simeq N^{- 3/4} (b/l_{\rm B}) ^{1/2}$ the 
electrostatic interactions are relevant and  the end-to-end distance
can be written in the form
\begin{equation}
R \simeq N f^{2/3} (l_{\rm B} b^2) (\log N)^{ \gamma } .
\label{scaling}
\end{equation}
In this equation  the value of $\gamma = 1/3$ is confirmed also by a
simplified chain under tension variational method 
\cite{Gennes2,Barrat} where the trial Hamiltonian
is the Hamiltonian of a Gaussian chain subject to an external tension
$ \vec{F}$. Indeed we also reproduce
this result for large enough values of the Coulomb strength 
as we will discuss in the next section. We
observe that the exponent $ \gamma $ is non universal, i.e. decreasing
with the interaction strength, and approaching the limiting value
$\gamma = 1/3$ from above in the region of $ \beta > 1$, in agreement
with equation (\ref{inequality2}). At $ \beta = 1$, the value
of $ \gamma \simeq 0.38$ is in good agreement with the calculation of
S\"oderberg and collaborators \cite{Sodeberg2}. It is
important to remark that if we assume  in the previous asymptotic
analysis that the  $ \gamma $ exponent is universal  a spurious nested
logarithmic form for the end-to-end distance (the way Allegra \cite{Allegra2}
discussed long time ago) is obtained. This scenario however is not confirmed by
the numerical analysis of the Euler equations as will be presented
in section III.
The balancing between the elastic energy and the interaction term
determines the exponents (\ref{inequality0})-(\ref{inequality})
for the self avoiding
chain (or short-range interaction) problem, whereas  for the Coulombic
case  the result is given in equations(\ref{inequality1})-
(\ref{inequality2}).

The term that appears on the l.h.s. of equation (\ref{asymptotic}) is
negligible
in this case. The proper condition for this to happen reads
$1+ 2 \nu >2$, which leads to the condition $d<4$ for the
self-avoiding chain \cite{Gennes} and $d <6$ for the Coulombic chain
\cite{Pfeuty}. Accordingly the upper critical dimensionality is
$d_c=4$ for the former case and $d_c=6$ for the latter one. At $d>d_c$
we can neglect the interaction term and the chain becomes Gaussian.

We would like to make a general remark at this point. It is clear
that the variational technique cannot cover the results of the 
RG. From the universality of the self avoiding walk problem it is well known
that the choice of the parameter $w$ does not change the 
scaling exponent and below the upper critical dimension no logarithmic
corrections are necessary. At the critical dimension, these become
obviously real. On the other hand the variational technique produce
these corrections naturally and works well in the mean field limit.
For the case of polyelectrolytes the situation is more complex.
The Coulomb potential changes its nature from long range to short range
between three and four dimensions. Therefore the fixed point and scaling
behavior is 
completely different,  
as it has been discussed already by 
de Gennes \cite{Gennes}. It is therefore also necessary to check our
results by alternative methods.

\section{Numerical Solution of the Euler equations}

In this section we consider in detail the Euler equations
(\ref{Euler5}) - (\ref{Euler2}) for a polymer chain of length $N$ with
cyclic boundary conditions, obtained by minimization of the
variational free energy $F_V$.
Two different ways in order to solve the variational principle
minimization for a finite chain of length $N$ are possible.
On the one side, a multidimensional minimization of the free energy $F_V$
in a $N$ dimensional space is considered \cite{Numrec}. On the other
side, a direct solution of the Euler equations will be presented.

We checked first that our multidimensional algorithm is able to
reproduce the
simple Gaussian profile for the simple case of the non interacting chain
problem. We then implement our algorithm by choosing
a very small value of the interaction strength $ \beta$, where the
solution is expected to be very close to the 
simple Gaussian chain solution, and give
$N+1$ trial profiles for the function $g_N(k)$. The multidimensional
algorithm  via a sequence of reflections and translations between the
$N+1$
trial profiles $g_N(k)$, relaxes to the true solution that can be
resolved
within the numerical precision desired. Once the algorithm was properly
tuned, we moved on studying systematically the shape of the optimal
profile that
minimizes  the free energy $F_V$ for increasing values of the
interaction
strength $\beta$.
For any specific value of  $ \beta$ the optimal profile was
checked to satisfy the basic symmetry requirements given in equation
(\ref{Euler3}).
Once the numerical solution $g_N(k)$ that minimize the variational free
energy
was found, we proceeded to analyze the properties of the solution in the
asymptotic regime, in order to check our
results for the maximal mean square distance $b(N/2)$. Results for the
mean square distance $\langle ({\vec r}_{i+n} - {\vec r}_{i})^2 \rangle$
for intermediate values of $n$  were also studied in order to confirm
the phenomenological predictions of the electrostatic blob picture.

A direct implementation of the equations (\ref{Euler5}) - 
(\ref{Euler2}) does confirm the 
multidimensional minimization of the free energy $F_V$ and also
represents a convenient method to solve the variational problem
itself. We also consider, for the case of $ \lambda =1$ a Fast 
Fourier Algorithm that let us consider chains of length up to 
$N = 2^{15}$. 
The difference between the values of the exponent
$\gamma$ computed with $N<1024$ and $N<32768$ are very small and tends 
to vanish when the interaction strength increases. We consider this as 
a second, independent test to confirm the non-universal behavior of $ \gamma$.
For this reason we will consider shorter chains only for what concerns 
the case of $ \lambda=2$ and for the Self Avoiding Chain Problem (see
section IIc).
We  start with a purely Gaussian profile in order to compute the
left hand side of equation (\ref{Euler2}), via the Fourier transform of
(\ref{Euler1}) that defines the mean square distance $b(n)=\langle
(\vec {r_{i+n}} - \vec {r_{i}})^2 \rangle$ for a small value of the
interaction strength $ \beta$ and iterate this procedure until
convergence
is achieved. Once the solution of the Euler equations is found for a
specific value of the interaction, one can use it as a starting point for the
following iteration procedure, in order to solve the 
same set of equations for a
higher value of $ \beta$. It is important to note that a simple recursive
implementation of the Euler equations above is highly unstable  in the
sense that the number of iterations required as a function of $\beta$ rapidly
increases, making a simple iteration procedure quite ineffective.
A simple solution to this problem can be found borrowing ideas from
the renormalization group literature \cite{berker}.
Renormalization Group Theory was successfully applied to determine the
phase diagram and magnetic properties of disordered spin systems
\cite{berkereio}. A functional set of recursion relations for the
probability distribution function of bond randomness was studied and a
fixed distribution was found for different values of the initial
conditions.
Instead of a simple iterative procedure, which could be a rather slow
converging process, a multidimensional bisection between different
initial conditions for the renormalization group trajectory turns out to
be a very effective method in order to achieve convergence.
In a similar fashion, instead of a simple iteration of the Euler
equations above, we prefer to bisect between possible initial profiles
of the function  $g(k)$ given above.

\subsection{Results for the Coulombic chain}

We obtained the solution of the Euler equations
(\ref{Euler5})-(\ref{Euler2})
via the iterative method discussed above and  independently  via the
direct minimization of the variational free energy $F_V$ in equation
(\ref{Euler4}) within the numerical precision .
We considered cyclic chains of length $2^4<N<2^{15}=32768$, for different
values of the interaction strength $\beta$ and two different forms of
the  potential (\ref{potential}), corresponding to the values
$ \lambda=1,2$.
We report the results of the end-to-end distance for different
values of $ \lambda$  and  for chains
of increasing size $N$  at values of $ \beta=0.5$ and $ \beta =1.0$.

\vspace{1.2cm}
\begin{tabular}{|c|c|c|c|c|c|c|c|c|c|c|c|} \hline
$ \lambda$ & $ \beta$  & N=120 & N=240 & N=320 & N=520 & N=1024&N=2048&N=4096&N=8192&N=16384&N=32768 \\ \hline
1.0  & 0.5 & 60.6 & 128.2 & 174.5 & 293.2 & 601.7 &1249.4&2583.4&5372.2&11071.2&22769.8 \\ \hline
1.0 & 1.0  & 78.0 & 164.4 & 223.6 & 374.2 & 767.59 &1590.6 &3283.9&6778.8&13942.4 &28627.3 \\ \hline
\end{tabular}
\begin{center}
\begin{tabular}{|c|c|c|c|c|c|c|} \hline
$ \lambda$  & $ \beta $ & N=120 & N=240 & N=320 & N=520 & N=1024 \\ \hline
2.0  & 0.5  & 26.2 & 48.6 & 63.0 & 98.12 & 140.2\\ \hline
2.0  & 1.0 &33.2 & 62.5 & 81.4 & 126.2 &248.7   \\ \hline
\end{tabular}
\end{center}
{\em Tab.1:} Results for the end-to-end distance $R$ for different 
values of the chain length  $N$ and  the interaction strength $\beta$ and
$\lambda$. These results are obtained solving explicitly the
the Euler equations (\ref{Euler5})-(\ref{Euler2}).

Let us begin to discuss the case of a simple polyelectrolyte chain in
$d=3$ with pure Coulombic interaction ($ \lambda=1$). We analyzed the
end-to-end distance behavior for increasing values of $N$,
i.e. $b(N/2)$. 
This is reported in Fig.1. Dependence of the exponent $ \gamma$  
on the interaction strength suggests a weak violation of universality 
as manifest in Fig.2.
\begin{figure}
  \begin{center}
    \includegraphics[scale=0.45]{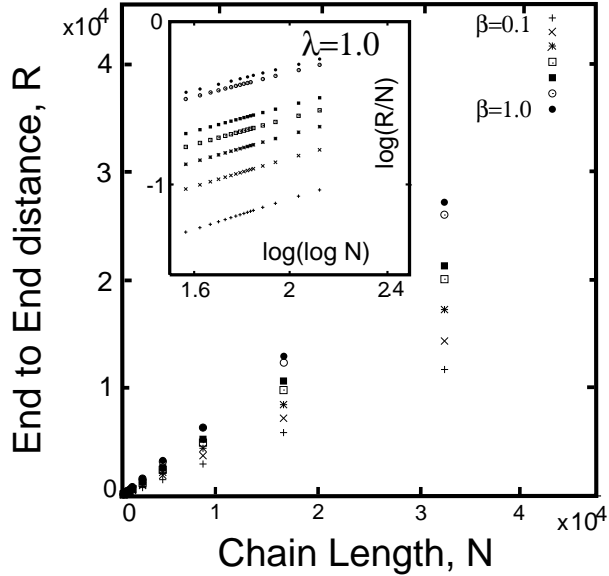}
  \end{center}
  \caption{End-to-End distance $R$ versus the chain length $N$, 
   for increasing values of the interaction strength $ \beta$.
    The inset shows the logarithmic plot, when a linear dependence
    of $R$ on the chain length is assumed, beside logarithmic corrections, for
    values of $\lambda=1$ in equation (\ref{potential}).}\label{fig1}
\end{figure}

\begin{figure}
  \begin{center}
    \includegraphics[scale=0.45]{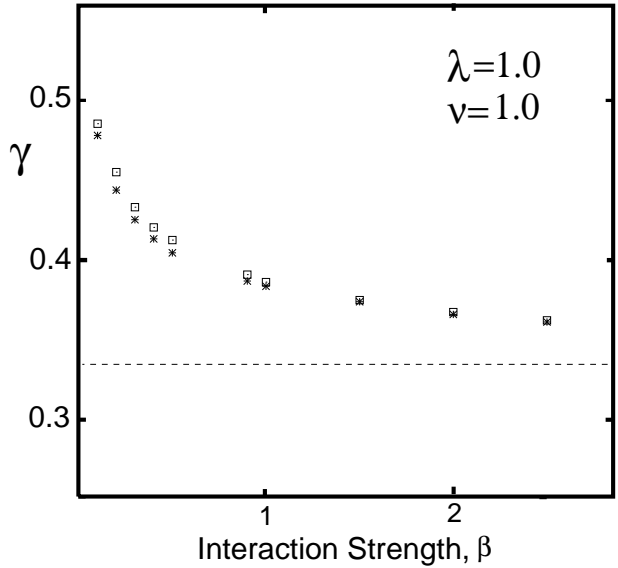}
  \end{center}
  \caption{The exponent $\gamma$ as measured (see Fig.1) for different
values of the interaction strength $ \beta$. The exponent $ \gamma$
is seen to decrease for large values of the interaction strength,
towards the expected value $\gamma=1/3$. The squares indicates the
computed value of $\gamma$ with chains of length $N<1024$ while stars
show the computed value of $\gamma$ for chains of length $N<2^{15}$.
 }\label{fig2}
\end{figure}

The des Cloizeaux argument, once extended to the
polyelectrolyte chain problem in $d=3$  implies an
overstretching ($\nu = 2/ \lambda =2$), while the
 ansatz suggested above
indicates $\nu = 3/(\lambda +2)=1$.
In order to compute the non-universal exponent $ \gamma$  we plot 
$R/N^{3/ (\lambda+2)}$  as shown in the inset of Fig.1. 
The non-universal behavior of $\gamma$ agrees with the 
inequality (\ref{inequality2})  derived in the asymptotic 
analysis of section II.
In Fig.2, the exponent $\gamma$ decreases with increasing
values of $ \beta$, approaching values very close to $ \gamma = 1/3$,
for large values of the interaction strength.
A recent discussion about the logarithmic correction to scaling
behavior for polyelectrolyte chains, has been given \cite{Sodeberg2} 
and a direct derivation of the value $ \gamma = 1/3$ has been
obtained. In the notation of reference
\cite{Sodeberg2} the small temperature limit $T \rightarrow 0$
corresponds to the large interaction strength limit $\beta
\rightarrow \infty$ which is  consistent  with our findings. The result
\begin{equation}
R \propto N (\log N)^{1/3}
\end{equation}
obtained in the large $ \beta \rightarrow \infty$ limit, can be
interpreted as a stretching of the harmonic bonds of the chain.
The value of $\gamma = 1/3$ was also obtained by a
simplified chain under tension variational method \cite{Barrat}
where the trial Hamiltonian is the Hamiltonian of a Gaussian chain
subject to an external tension $ \vec{F}$.
This result, confirmed also by the Monte Carlo simulations presented in
the next section, holds in the limit of large interaction strength but
should not considered as a limiting case of little practical impact.
Ordinary temperatures in the  aqueous solutions correspond to
surprisingly ``high'' values of the interaction strength $\beta$
\cite{Sodeberg2}. 

A rather important problem discussed in the past \cite{Bouchaud}
is the polyelectrolyte chain with an interaction (\ref{potential}) 
at  $ \lambda=2$. After the proper
rescaling of the interaction strength $\beta$ (see equation
\ref{Euler4}), this problem can be thought as a chain with pure Coulombic
interaction for  $d=4$ . 
Results for the end-to-end distance versus the chain length $N$
are shown in Fig.3.
\begin{figure}
 \begin{center}
 \includegraphics[scale=0.4]{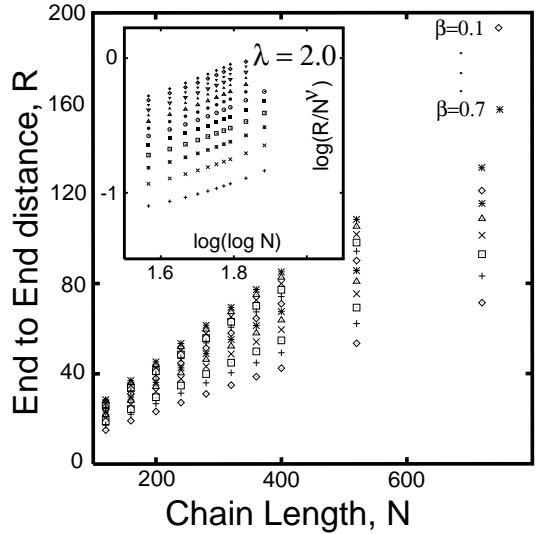}
 \end{center}
 \caption{End-to-End distance $R$ versus the chain length $N$
   for increasing values of the interaction strength $ \beta$ at $ \lambda=2$ .
    The inset shows the logarithmic plot, when a power law 
    $N^{ \nu_F}$ dependence
    of $R$ on the chain length is assumed, beside logarithmic 
    corrections.}
 \label{fig3}
\end{figure}
A similar discussion about the non-universal
behavior of the exponent $\gamma$, follows the same lines as in
the previous case (Fig.1-Fig.2). According to the  ansatz we proposed in the asymptotic analysis
of section II C   we compute the exponent
$\gamma$ as in Fig.4, where $ \nu= \nu_F$ is the Flory exponent.
\begin{figure}
  \begin{center}
    \includegraphics[scale=0.45]{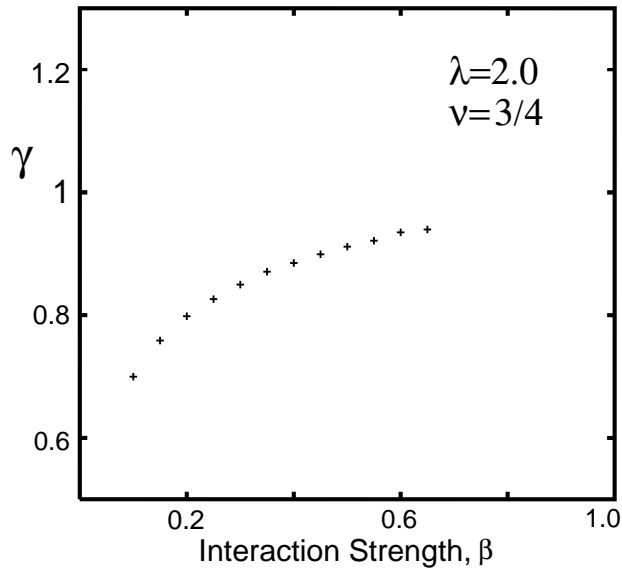}
  \end{center}
  \caption{The exponent $\gamma$ as measured from Fig.3 for different
   values of the interaction strength $ \beta$. The exponent $ \gamma$
   is seen to increase for large values of the interaction strength
   towards the value $\gamma \approx 1.0$.}\label{fig4}
\end{figure}

In Fig.5 we compute the non-universal behavior of the exponent
$\gamma$ assuming a linear dependence of the end-to-end distance
of the chain length $N$, beside logarithmic terms. This is not the
case, for the potential (\ref{potential}) with
$\lambda=2$, but interesting comparison with previous results follows.
The exponent $ \gamma$, as in Fig.5, approaches values of
$\gamma \approx - 0.4$ at values of $ \beta =1.0$ where Monte Carlo
calculations were performed \cite{MaPa} and where the computed value
of $\gamma \approx -0.35$ is in good agreement with our result. 
A two parameter fit would
represent a very conclusive way to discriminate between these two
possibilities, but the chain lengths that one can reach in 
simulations and in a direct minimization of the variational free
energy  are not long enough to make such analysis effective.

\begin{figure}
  \begin{center}
    \includegraphics[scale=0.45]{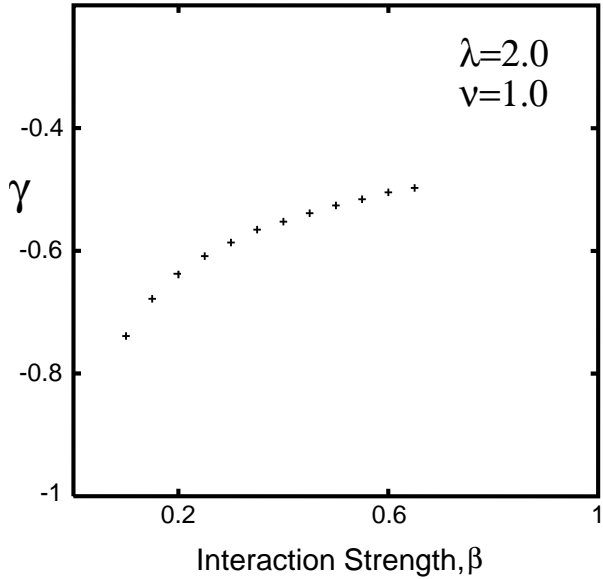}
  \end{center}
  \caption{The exponent $\gamma$ as measured from Fig.3 for different
    values of the interaction strength $ \beta$. The exponent $ \gamma$
    is seen to increase for large values of the interaction strength
    towards the value $\gamma \approx 1.0$, once a linear dependence
    of $R$ on the chain length is assumed, beside logarithmic 
    corrections, for values of $\lambda=2$ in equation
    (\ref{potential}). This results should be compared with the 
    value of $\gamma \approx -0.35$ previously found via Monte Carlo 
    simulations, for values of $\beta=1.0$.}\label{fig5}
\end{figure}

\subsection{Electrostatic blob picture}

We will now present a numerical analysis of the
electrostatic blob picture. We measured the crossover (shown by 
the mean square distance as a function of the polymer chain length
$N$) from a Gaussian regime, observed at small distances, to the
stretched regime  for long enough distances. This represents an
effective way to compute the electrostatic blob size and to check
our method. 
One of the major achievement of the Gaussian variational approach
discussed in section I is that it does not simply provides a quantitative
description of the large $n$ behavior 
correlation function. A full numerical solution of the Euler
equations (\ref{Euler5})-(\ref{Euler2}) provides informations on the
intermediate behavior  of the correlation functions
$b(n)= \langle (\vec {r_{i+n}} - \vec {r_{i}})^2 \rangle$.
That is why it is interesting to check whether short enough 
segments of the polymer chains behave  as Gaussian chains, as 
expected from general scaling arguments \cite{Gennes2},  
whereas  the large $n$ behavior corresponds to  equation (\ref{scaling}).
Once the crossover is observed,
the size of an electrostatic blob $ \xi$ can be extrapolated. The
dependence of the electrostatic blob size $ \xi$ from the interaction 
strength $ \beta$, that we expect from phenomenological arguments, 
is indeed confirmed within our variational approach (see Fig.6).
In particular we expect
\begin{eqnarray}
 R & \propto & (N/g) \xi \simeq N a \beta ^{1/3} \nonumber  \\
 \xi & \propto & a \beta ^{-1/3}
\end{eqnarray}

In Fig. 6 we show the electrostatic blob size as a function of the
interaction strength $\beta$. The $ \xi \propto \beta^{-1/3}$ behavior 
expected from standard scaling arguments is here also found.
\begin{figure}
  \begin{center}
    \includegraphics[scale=0.4]{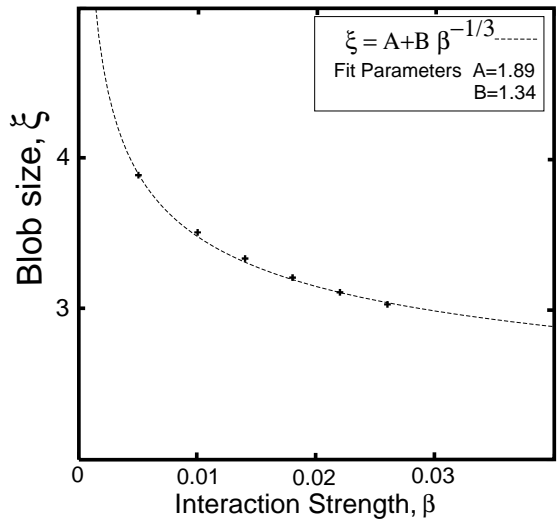}
  \end{center}
  \caption{Blob size, as measured from the direct solution of
  equations (\ref{Euler2})-(\ref{Euler4}), considering the crossover
   between the small distance behavior of the mean square distance
   $b(n)$ versus the large distance behavior.  Different values of the
   electrostatic blob size for increasing values of the interaction
   strength $\beta$ are compared with the $\xi \approx \beta^{-1/3}$
   behavior expected from phenomenological arguments. }\label{fig6}
\end{figure}

\subsection{Results for the Self Avoiding Chain problem}

We consider the problem of a polymer chain with a short
range interaction between the monomers, in a good solvent.
Within the Gaussian Variational Principle,
the expression for the free energy and the Euler equation are given by
equations (\ref{Euler4bis})  and (\ref{Euler2bis}). In section II we have
discussed the dimensional shift (\ref{shift}).
Nevertheless this equivalence is purely formal and no deep insight
is hidden behind it. The two problems considered are indeed pretty
different and
they reduce to the same type of variational equations because of the
Gaussian approximation itself.
It is important to remark that the Gaussian Variational
Principle, specially reliable for the long
ranged polyelectrolyte problem, has been originally introduced to study
the self avoiding homopolymer chain problem
\cite{Cl,Cloizeaux}. Even though it is nowadays well established that
the Gaussian Variational Principle is not very well suited for the short
ranged interacting chain  problem, we wish to make clear that
within this approximation, apart for spurious logarithmic corrections,
the usual Flory-like power law behavior should be recovered.

In the original asymptotic analysis of des Cloizeaux the
non integer value of $ \delta$ is tacitly assumed, so that the result
of $ \nu = 2/d$ for the swelling exponent was obtained. Although
we believe this result to be incorrect, it is often mentioned in the literature
to discuss the asymptotic properties of polyelectrolyte chains
\cite{NO}.
According to our analysis and to the previous analysis of Allegra and
coworkers \cite{Allegra1},  the problem of a self avoiding chain in good
solvent turns out, within the Gaussian Variational Principle, to present
the expected Flory-like behavior. This is not surprising since the original
derivation of Flory  also belong to the class of Gaussian Variational 
approximations, since a Gaussian probability distribution of the
distance between the ends of the chain is assumed. A numerical
justification of the des Cloizeaux result for the swelling exponent
can also be found in the literature \cite{Bratko}. 
We believe that this numerical analysis is inconsistent since 
the constraint on the interaction 
parameter (assumed by des Cloizeaux and critically reviewed in section
II)  is enforced in the numerical algorithm. This makes the
results obtained highly questionable. In Fig. 7 and Fig. 8 we present
the results of the end-to-end distance for the self avoiding homopolymer
chain problem studied via the numerical solution of the Euler equations
discussed above. 
\begin{figure}
  \begin{center}
    \includegraphics[scale=0.45]{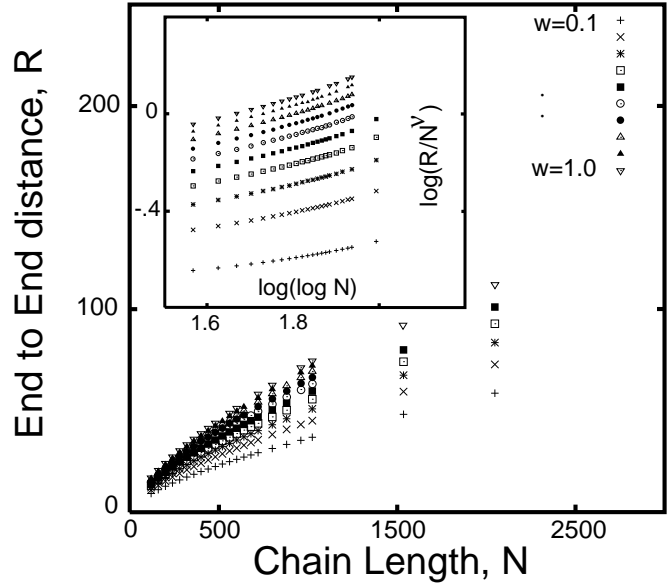}
  \end{center}
  \caption{End-to-End distance $R$ versus the chain length $N$.
   for increasing values of the interaction strength $w$, that
   represent in this case the quality of the solvent.
    The inset shows the logarithmic plot, when a $N^{ \nu_F}$ dependence
    of $R$ on $N$ is assumed, beside logarithmic corrections. }
 \label{fig7}
\end{figure}
Results for the end-to-end distance as a function of 
the polymer chain length $N$ are shown and a quantitative analysis of
the non universal logarithmic part  to be considered as a spurious 
effect due to the short range nature of the problem is presented. 
\begin{figure}
  \begin{center}
    \includegraphics[scale=0.45]{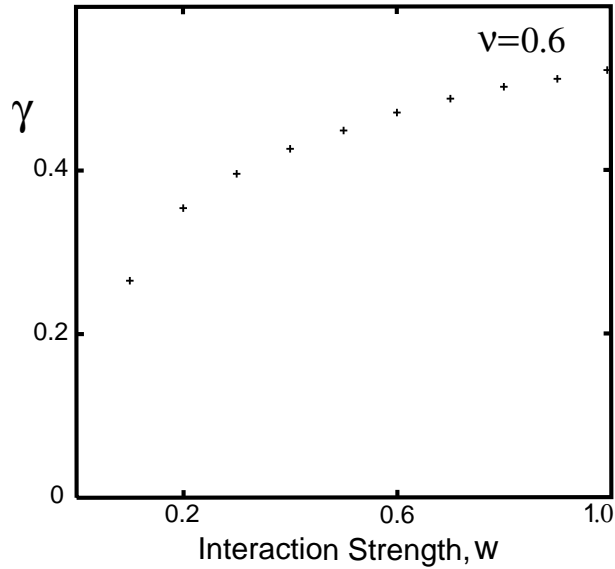}
  \end{center}
  \caption{The exponent $\gamma$ as measured from Fig.7 for different
values of the interaction strength $w$. The exponent $ \gamma$
is seen to increase for large values of the interaction strength.}
\label{fig8}
\end{figure}
According to the asymptotic analysis given above   we expect the
exponent $\gamma$ to be non universal and to depend indeed on the 
quality of the solvent.

\section{Numerical Simulations}

In  order to test the conclusions obtained in the previous section we
decided to perform a numerical simulation for a polyelectrolyte chain in
an ideal salt free solution, when pure Coulombic interactions are
considered between the monomers. 
A bit of caution is required  since we
will compare the results of our simulations with the results obtained
via the Gaussian variational principle, where a cyclic geometry for the
polymer chain is chosen.
It was easy however to device a numerical simulation of the
Monte Carlo type where close chain configurations can be taken under 
investigation.
As remarked long time ago, \cite{Cloizeaux} the cyclic geometry chosen
within the variational calculation becomes unessential as soon as long enough
chains are considered and this turns out to be confirmed by our numerical
simulations.
We then consider a Monte Carlo simulation, performed in
the canonical ensemble via the standard Metropolis algorithm.
We consider a pivot move for what concerns the elementary Monte Carlo
move, i.e. we assume rigid bonds between the monomers and an
elementary pivot move consists in a random rigid rotation of a
randomly chosen part of the chain. The
pivot algorithm, first described in reference \cite{Sokal} and lately
applied in the context of polyelectrolyte chains with and without
salt \cite{Sodeberg2} is a very effective method to confirm our
general conclusions.
Several million Monte Carlo moves (the number  which was properly increased with the
chain length) are required in order to reduce below
$1 \% $ percent the statistical fluctuations for the end-to-end
distance. We also combine our Monte Carlo simulation with a 
Molecular dynamics subroutine that is randomly called, during the Monte Carlo 
simulation, with a certain fractional occurrence ratio. On the one side,
the pivot algorithm is well suited to simulate the polymer chain at
long enough scales. The pivot moves are in a sense collective moves
and 
a good description of the chain is certainly obtained at long scales.
On the other side the Molecular Dynamic algorithm is well suited to
simulate the chain at small distances. 
We initially perform independent simulations with the Monte Carlo 
and the Molecular dynamics algorithm in order to tune the simulation 
parameters involved  for a chain of a given length $N$.
Once the two methods were independently tested, we proceeded to combine 
them as discussed above.

We now present our results for the end-to-end distance for 
$d=3$  polyelectrolyte problem. In Fig. 9 the end-to-end
distance is shown as a function of the polymer length $N$ for the 
specific value of $ \beta =1.0$. 
The inset shows the end-to-end distance divided by the pure 
power law Flory-like dependence $N^{\nu}$ as a function of $ \log N$. 
We repeated the same analysis of  Fig.9 changing the value of the 
interaction strength $ \beta$.
The non universal behavior obtained in the 
asymptotic analysis of section II and in the numerical analysis 
of the Euler equations (\ref{Euler5}) - (\ref{Euler2}), is found 
within our simulation results (see Fig.10).

\begin{figure}
  \begin{center}
    \includegraphics[scale=0.45]{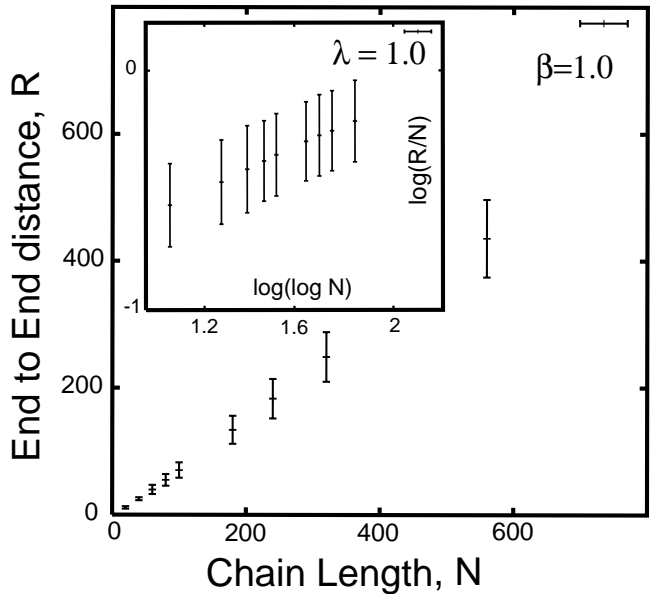}
  \end{center}
  \caption{End-to-End distance $R$ versus the chain length $N$
   for the  polyelectrolyte chain at increasing values of the
   interaction strength $ \beta$ and  $d=3$, obtained via the pivot
   Monte Carlo algorithm combined with Molecular Dynamics for the 
   value of the occurrence factor $r=0.1$. 
   Different simulations, corresponding to 
   different values of the occurrence factor $r$ have been shown to be
   numerically consistent within each other. The inset shows the 
   logarithmic plot, when a linear dependence of $R$ on $N$ is 
   assumed, beside logarithmic correction}\label{fig9}
\end{figure}

The non universal exponent $ \gamma$ is seen to decrease, as we 
obtained with the variational method of section II-III. Nevertheless one
should treat  the results of simulation with  a bit caution. 
\begin{figure}
  \begin{center}
    \includegraphics[scale=0.45]{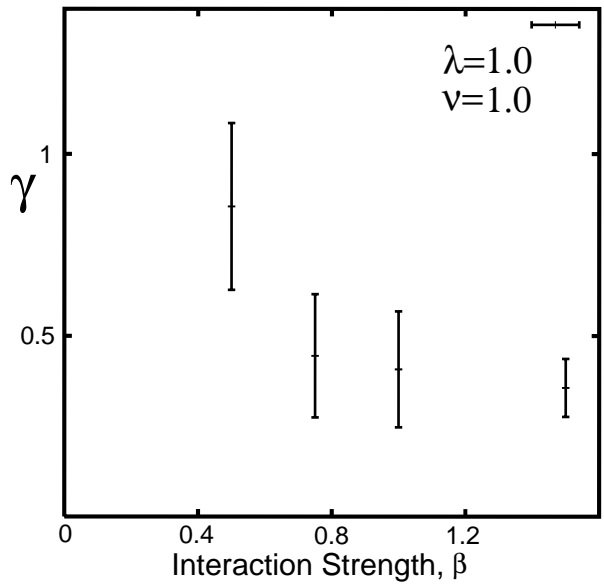}
  \end{center}
  \caption{The non universal exponent $ \gamma$, for different
   values of the interaction strength $\beta$ as measured from
   Fig.9, for the specific value of $ \beta=1.0$ and for the
   values $ \beta=0.5,0.75,1.5$.}
\label{fig10}
\end{figure}
In the numerical simulations monomers are
considered to be rigidly connected  while in the variational principle
approach monomers are connected by springs. In this case one can not
expect  that the exponent $ \gamma$  as a function of interaction 
strength has the same large $ \beta$ behavior. 

\section{Conclusions}

In the present paper we have reinvestigated the Gaussian variational principle 
and its applicability to polymer chain with long
and short range interactions. 
The asymptotic analysis originally presented by des
Cloizeaux in the context of self avoiding walks
has been critically reconsidered. In particularly we have shown that
the des Cloizeaux's analysis overconstains the model's parameters. It
has been found that within the Flory type of balancing conditions 
(when the elastic term is balanced by the interaction but the
entropic term is not relevant) the asymptotic law (\ref{newansatz})
satisfies the Euler equation (\ref{Euler2}) . In this asymptotic law 
the Flory exponent $\nu = 3/(d+2)$ and logarithmic
corrections to
scaling  is characterized by a non - universal exponent $\gamma$. 
Within the Gaussian variational principle the universal
exponents for the case of the Coulombic and short range interactions
are related by a dimensional shift (\ref{shift}). 

Gaussian variational principle suits much better for the long range
problems which was already discussed in reference \cite{Bouchaud}. The
arguments we have presented prove that the scaling form
(\ref{newansatz}) is valid for a general long - range interaction
potential (\ref{potential}). The exponents $\nu$ and $\gamma$
satisfy the equations (\ref{florexp}) - (\ref{inequality2}).
According to the condition (\ref{inequality2}), the 
exponent $\gamma$ is non - universal and depends from the interaction
parameter as well as from the chain model (compare e.g. Fig.2 and
Fig. 10) . In this sense we may  consider the combination $l_{\rm eff} =
a (\log N)^{\gamma}$ as an effective Kuhn  segment for chains with
unscreened type of interactions.

On the contrary, for the short range interactions the Gaussian
variational principle simply overestimates the probability of the
segment - to - segment contacts \cite{Bouchaud} which leads to more
extended configurations compare to the Monte Carlo results
\cite{Sokal}.
We conclude observing that the models which are characterized by
the interplay between long - ranged attraction and repulsion are also
useful systems to be treated by the Gaussian variational
principle, as presented here. Appropriate systems
are polyampholyte \cite{Higgs}. For the randomly charged
polyampholytes Kantor and Kardar \cite{Kantor} have found strongly
nonuniversal behavior. For example at $d < 4$ the spatial
configurations depend on the polyampholyte excess charge $Q$ so that
if $Q > Q^* \approx e\sqrt{N}$ the compact form changes to the
stretched one. We believe that the Gaussian variational principle is a 
good starting point for the investigation of such systems.

\section*{Appendix A}

The asymptotic analysis of equations (\ref{Euler2bis}) 
suggests to consider the following series
\begin{equation}
S( \delta ,z) = \sum_{n=1}^{\infty}
\frac{ z^{ n}}{n^{ \delta}}  ,
\label{series1}
\end{equation}
In the asymptotic limit $ k \rightarrow 0$ the interaction term
of equation (\ref{Euler2bis}) is simply related to $S ( \delta, z)$. 
  Even the simple power law ansatz $b(n) \propto n^{ \nu}$
requires  some care. The expression ( \ref{series1}) is evaluated exactly but 
two rather different expressions are obtained if the exponent 
$\delta=\nu (d+2)$ is assumed to be integer or non-integer. 
The case of the  exponent $\delta $ being non-integer reduces 
the interaction term to the expression given in equation (\ref{series})
and leads to the des Cloizeaux result. 
Assuming an integer value for the  exponent $\delta $  
produces instead a different expression for the interaction term 
and the proper balancing condition with the elastic energy naturally 
follows. 

Let us consider the series $S( \delta,z)$ in this two distinct 
cases. 
In equation (\ref{series1}) $z$ and $ \delta$ can be complex quantities, as
soon as $|z| \le 1$ and $Re~ \delta >1$ at $|z|=1$ whereas  $Re~\delta >0$ if
$|z|<1$.   The calculation of the series  (\ref{series1}) ( see sec. 1.11 in
ref.\cite{erdelyi} )
 gives 

\begin{equation}
S( \delta,z)= \Gamma( 1 - \delta) [ \log( \frac{1}{z} ) ]^{ \delta -1} +
\sum_{r=0}^{ \infty} \zeta ( \delta -r) \frac{ ( \log z)^r}{r!},
\label{seriesexp}
\end{equation}
where $| \log z | < 2 \pi$, $\Gamma(x)$ is the gamma function,
$ \delta \ne 1,2,..$ and $ \zeta(x)$ is the $\zeta$-zeta
Riemann's function.
We are interested in the case of $z= \exp(ik)$ at $k \rightarrow 0$.
Considering the real part of both sides of equation (\ref{series0})
in the asymptotic limit $ k \rightarrow 0$, we obtain equation
(\ref{series}).

We now consider the case of integer values of $\delta$. The $\Gamma
(x)$ has poles at all negative integer arguments whereas the pole of
$\zeta(x)$  is placed at $x = 1$.  We write $ \delta =m + \epsilon$ where $m$ is a positive
integer and $ \epsilon \rightarrow 0$. Then in the vicinity of poles  the gamma and
$\zeta$ - 
functions  can be rewritten as 
\begin{eqnarray}
\Gamma(1 - m - \epsilon ) &=& \frac{(-1)^m}{(m-1)!}  \left\{ \frac{1}{
    \epsilon} - \psi(m) + {\cal O}(\epsilon)\right\}\nonumber\\
\zeta( 1 + \epsilon ) &=& \left\{ \frac{1}{ \epsilon} - \psi(1) +
  {\cal O}(\epsilon) \right\}.
\label{asym3}
\end{eqnarray}
We should also take into account  that
\begin{eqnarray}
\left[\log \left(\frac{1}{z}\right)\right]^{\epsilon} = 1 + \epsilon
\log \log \left(\frac{1}{z}\right) + {\cal O} (\epsilon ^2).
\label{asym4}
\end{eqnarray}
After that the equation (\ref{seriesexp}), due to the cancellation of poles in
gamma and $\zeta$ - functions at small values of
$\epsilon$, becomes \cite{erdelyi}
\begin{eqnarray}
S(m,z) &=& \frac{ ( \log z )^{m-1} }{(m-1)!}
\left( \psi(m)-\psi(1) - \log \log (1/z) \right) \nonumber \\
&+& \mathop{{\sum}'}_{r =0 }^{ \infty }  \zeta( m-r) \frac{ ( \log z)^r
}{r!},
\label{asym}
\end{eqnarray}
where the prime indicates that the term $r=m-1$ is to be omitted.
In equations (\ref{asym}) (\ref{asym3}) $ \psi(x)$ is the digamma function,
defined as the logarithmic derivative of the gamma function
$\psi( \delta) = \frac{ d \log \Gamma( \delta)}{d \delta}$.
In the limit $ k \rightarrow 0$, for $z = \exp(ik)$ and $ \delta=3$,
eq. (\ref{asym}) have been considered in eq.(\ref{asym2}).

We want to show now that the integral representation of the sum is correct in the 
asymptotic limit and that finite $n<l$ terms do not contribute in this 
limit. Let us consider explicitly the presence of a finite 
cutoff in the series (\ref{series1}). 
\begin{equation}
S( \delta ,z) = \sum_{n=l}^{\infty}
\frac{ z^{ n}}{n^{ \delta}}  ,
\label{series2}
\end{equation}
which reduces to equation (\ref{series1}) at $l=1$. For integer values
of $\delta$ it is easy to derive exactly the following expression \cite{erdelyi}:
\begin{eqnarray}
S(m,z,l) &=& \frac{ ( \log z )^{m-1} }{(m-1)!}
\left( \psi(m)-\psi(l) - \log \log (1/z) \right) \nonumber \\
&+& \mathop{{\sum}'}_{r =0 }^{ \infty }  \zeta( m-r) \frac{ ( \log z)^r
}{r!}.
\label{asymm2}
\end{eqnarray}
Consider the series (\ref{series2}) with a finite cutoff $l$ in the 
asymptotic limit $k \rightarrow 0$. We want to evaluate the 
contributions to the series from finite $n<l$ terms, when 
 $m=3$. We obtain
\begin{equation}
\sum_{n=l}^{\infty}
\frac{1-\cos(nk)}{ n^{3} }  \simeq \frac{k^2}{2} ( \psi(3)- \psi(l) -\log
k ) +O(k^4),
\label{series3}
\end{equation}
so that the contribution from finite $n$ takes the following form
\begin{equation}
\sum_{n=1}^{l}
\frac{1-\cos(nk)}{n^{3}} \simeq \frac{k^2}{2} ( \psi(l)- \psi(1) )
\label{series4}
\end{equation}
Summing up equations (\ref{series3}) and (\ref{series4}) we obtain
again the result (\ref{asym2}). The expression
(\ref{series3}) shows that for small $k$ values only
large $n$ terms contribute to the sum (\ref{series2}).
Indeed the contribution coming from $n>l$ produces a term $k^2 \log(k)$.
This, whenever $k << \exp(- \psi(l)) $, will dominate the simple $k^2$
behavior. Moreover, for large values of $l$ (see ref. \cite{erdelyi}) we have
$ \psi(l) \propto \log(l)$ and the condition above takes the form
$k << 1/l$. It follows that for $k << 1/l$
the term $k^2 \log(k)$ dominates and the sum does not depends on the
low limit value. In this case, the series can be well approximated
by an integral, i.e.

\begin{equation}
S(k) =\sum_{n=l}^{ \infty} \frac{1-\cos(nk)}{n^3} \simeq
\int_l^{ \infty} dn  \frac{1-\cos(nk)}{n^3} =
k^2 \int_{kl}^{ 2 \pi} dx \frac{1-\cos(x)}{x^3} =k^2 I(k)
\end{equation}

In order to compute $I(k)$, consider the derivative of the previous
expression, $I'(kl) =-1/2kl$ so that $I(kl) \simeq - \frac{1}{2} \log(kl)+c$,
where $c$ is a constant, and $S(k) =\frac{k^2}{2} ( c- \log(kl)) +O(k^4)$.
Comparison with equation (\ref{series3}) shows that for $k << 1/l$
only large $n$ contributions are relevant.

\section*{Acknowledgments}

We thank Christian Holm, Burkhard D\"unweg and Ralf Everaers for 
continuous and fruitful discussions. In particular we thank Dr. Holm 
for several comments on the simulation part.
One of us (G.M.) wish to express many thanks to Giorgio Parisi and
Roland Netz.

\end{document}